# Alloy-like behaviour of the thermal conductivity of non-symmetric superlattices

E. Chavez-Ángel[1,*], P. Komar[1,2,†] and G. Jakob[1,2]

[1] Institut für Physik, Johannes Gutenberg Universität Mainz, Staudingerweg 7, 55128 Mainz, Germany.
[2] Graduate School Materials Science in Mainz, Staudingerweg 9, 55128 Mainz, Germany

E-mail corresponding author: emigdio.chavez@icn2.cat

**Abstract**: In this work, we show a phenomenological alloy-like fit of the thermal conductivity of $(A)_{d1}:(B)_{d2}$ superlattices with $d_1 \neq d_2$, i.e. non-symmetric structure. The presented method is a generalization of the Norbury rule of the summation of thermal resistivities in alloy compounds. Namely, we show that this approach can be also extended to describe the thermal properties of crystalline and ordered-system composed by two or more elements, and, has a potentially much wider application range. Using this approximation we estimate that the interface thermal resistance depends on the period and the ratio of materials that form the superlattice structure.

**Nomenclature**

| | | | |
|---|---|---|---|
| $A$ | Bulk property of the element A of the $A_xB_{(1-x)}$ alloy system, or layer A of A:B superlattice | $uc$ | Unit cell |
| $AB$ | Effective alloy property | $TBR$ | Thermal boundary resistance |
| $AMM$ | Acoustic mismatch model | $x$ | Digital alloy composition |
| $B$ | Bulk property of the element B of the $A_xB_{(1-x)}$ alloy system, or layer B of A:B superlattice | **Greek symbols** | |
| | | $\eta$ | Interface density of a superlattice |
| $C$ | Fitting parameter | $\kappa$ | Thermal conductivity |
| $DMM$ | Diffusive mismatch model | $\rho$ | Thermal resistivity |
| $d_i$ | Thickness of the layer i of A:B superlattice | **Subscripts** | |
| $L$ | Period thickness | $i$ | Element $i$ compound system |
| $MD$ | Molecular dynamics | $di$ | Thickness of the element $i$ of a compound system |
| $m$ | Number of unit cells | $bulk$ | Bulk property |
| $N$ | Total number of periods of a superlattice | $max$ | Maximum of the thermal conductivity |
| $NSSLs$ | Non-symmetrical superlattices | $SLs$ | Superlattices |
| $n$ | Ratio of material composition | $x$ | Digital alloy composition |
| $R_{12}$ | Thermal boundary resistance | | |
| $SLs$ | Superlattices | | |

**Current address:**
[*] Catalan Institute of Nanoscience and Nanotechnology (ICN2), CSIC and BIST, Campus UAB, Bellaterra, 08193 Barcelona, Spain.

[†] Photonics Group, Institute of Physics, Lodz University of Technology, Wólczańska 219, 90-924 Łódź, Poland.

**Introduction**

A deep understanding of heat propagation at the nanoscale and the tuning of the thermal properties of materials are topics of continuous research activities motivated from the increasing importance of thermal management and energy efficiency. In this sense, the thermal boundary resistance (TBR, units $W^{-1}m^2K$), which describes the impact of the junction between two materials on the heat flow, is a concept that is far from being well understood yet at nanoscale, where the surface and interfaces play a fundamental role in the determination of transport properties. Commonly, the impact of the interface on the thermal resistance is described by two approaches [1]: the diffusive mismatch model (DMM) and the acoustic mismatch model (AMM). Both models assume that the phonons are the main heat carriers and their propagation would only be possible if there is an overlap in the phonon density of states between the materials. The main difference between these models is related to the treatment of the phonon scattering at the interface. Whereas, the DMM assumes that the interaction of the phonons at the interface is completely diffusive and "the phonon-memory" is lost, i.e. the wavevector of the transmitted phonon is random and independent on the incident phonon at the interface. The AMM considers both a specular reflection of the phonons at a perfect interface and an elastic propagation across it. However, the validity of one model over the other is still not well-established. In general, they are considered as the lower and the upper limits for phonon transport through an interface. At high temperatures, the wavelength of the thermal phonons (the main heat carriers) is very small, then, the interfaces seem rough and the thermal transport is dominated by DMM. On the other hand, at low temperature where the wavelength of the thermal phonons is larger and the interfaces appear smooth and perfect, the AMM dominates. Particular attention must be paid to the case of superlattices (SLs) because the models mentioned above describe the behaviour of the single interfaces between two bulk materials.

For the case of the SLs, the experimental estimation of the interface TBR is commonly determined by applying pure diffusive thermal circuit model. In such a model, the thermal resistivity of the SL, defined as the reverse of the thermal conductivity ($\rho_{SL} = 1/\kappa_{SL}$), is treated as a superposition of the thermal resistivities of the individual layers plus the TBR of the interfaces ($R_{12} + R_{21} \approx 2R_{12}$), and it is given by [1]:

$$\rho_{SL} = \frac{1}{L}\left(d_1\rho_1 + d_2\rho_2 + 2R_{12}\right) \tag{1}$$

where $d_1$ ($d_2$) and $\rho_1$ ($\rho_2$) are the thickness and the thermal resistivity of the layer 1 (2), $L$ is the period thickness, $L = d_1 + d_2$, and $R_{12}$ is the TBR of a single interface. One can show that for symmetrical SLs, i.e. $d_1 = d_2$, a single averaged value of the TBR can be obtained from the slope of the thermal resistivity versus interface density, $\eta = 1/L$, i.e.:

$$\rho_{SL} = \frac{1}{2}(\rho_1 + \rho_2) + 2R_{12}\eta \tag{2}$$

For the case of non-symmetric superlattices, $d_1 \neq d_2$, the TBR can be estimated from the their harmonic mean, given by [2,3]:

$$R_{12} = \frac{1}{2}(L\rho_{SL} - d_1\rho_{1,bulk} - d_2\rho_{2,bulk}) \tag{3}$$

where $\rho_{1,bulk}$ and $\rho_{2,bulk}$ are the bulk thermal resistivities of the layer 1 and 2, respectively, and $\rho_{SL}$ is the inverse of the measured thermal conductivity of the SL. However, the strong dependence of the TBR on the roughness, disorder, dislocations, bonding and the intermixing of the materials at the interface [4] leads just to a rough estimation of the real values expected for a perfect interface.

In addition, experimental reports have shown [3,5–8] that the thermal transport in SLs may exhibit a crossover between coherent and incoherent transport along the layering axis. This depends on the $L$ and the coherence-length of the acoustic phonons. The change of the transport behaviour occurs when the $\eta$ is large enough to limit the propagation of all high frequency phonons (with a short wavelength and particle-like behaviour), then, the thermal transport is governed by the low frequency phonons (with long wavelength and wave-like behaviour). The transition between coherent-incoherent (wave-particle) transport is observed as a minimum in the thermal conductivity as a function of $L$ [7,9]. This effect comes from the competition between the phonons diffusively scattered by each interface and the band-folded ones. Therefore, the TBR will not be constant and will depend strongly also on the $L$. While it is natural to think that there should exist one unique value for the TBR of the heat flow from material $A$ to material $B$, in SLs this wave interference modifies the phonon velocities and the density of states and also creates forbidden energy bandgaps for phonons of certain energies. Then, in the coherent regime the wave interference will control the heat flow, impacting in the effective value of the TBR.

Thus one cannot keep the idea of thermal conductivities of an individual layer in a strict sense. However, it is often convenient to argue with the material parameters and a TBR that depends on the structure of the SL. In the following, we show that this idea has similarities to the thermal resistivity of alloy compounds. Based on the Norbury's rule of the thermal resistivity in alloy compound [10,11], we propose an analytical method to describe the thermal conductivity and TBR of the non-symmetric SLs (NSSLs). Our aim is to provide a general view of the fundamental aspects of the thermal transport and to give an estimation of TBR in such systems. Moreover we demonstrate that the Norbury's rule captures the essential ingredients for properties of two interacting systems.

**Results and discussion.**

The linear or quadratic interpolation of material properties, in general, is a very useful and easy way to estimate many physical parameters of the alloys. In the case of $Si_{(1-x)}Ge_x$, the bulk modulus, the linear thermal expansion coefficient, Debye temperature, specific heat, Raman modes, to name a few of them, can by described as [12]:

$$AB = xA + (1-x)B \qquad (4)$$

where $AB$ is the effective alloy property, $x$ is the alloy content, $A$ and $B$ are the bulk properties of the constituent elements of the system. Whereas, other properties such as: the thermal conductivity, lattice constant, density, melting point, infrared refractive index, among others obey a quadratic interpolation given by the Norbury's rule [12,13]:

$$AB = xA + (1-x)B + x(1-x)C \qquad (5)$$

where $C$ takes into account the contribution from the lattice disorder generated in the alloy due to the random distribution of the atoms in a sublattice. This term is usually called a bowing or nonlinear parameter. Based on the work of Abeles [14], Adachi [10,11] demonstrated that the thermal resistivity of several III-V ternary and group-IV alloys of type $A_xB_{(1-x)}$ can be easily described by using Eq. (5). By considering an alloy-like system with "$d_1$" and "$d_2$" amounts of the element $A$ and $B$, respectively, the whole quantity of material is given by: $L = d_1 + d_2$. Then, if we keep constant the total amount of material, the alloy content can be expressed as $x = d_1/L$ and the Eq. (5) is rewritten as:

$$\begin{aligned} AB &= \frac{d_1}{L}A + \left(1 - \frac{d_1}{L}\right)B + \frac{d_1}{L}\left(1 - \frac{d_1}{L}\right)C \\ &= \frac{1}{L}\left(d_1 A + d_2 B + \frac{d_1 d_2}{d_1 + d_2}C\right) \end{aligned} \qquad (6)$$

On first thought, it is remarkable the similarities between Eqs. (1) and (6). Now, if we assume that both expressions are identical, the TBR can be expressed as:

$$R_{12} = \frac{1}{2}\frac{d_1 d_2}{d_1 + d_2}C \qquad (7)$$

where $C$ is a fitting constant with thermal resistivity units (i.e., $W^{-1}$ m K). It is important to notice the similarity of Eq. (7) with the description of the electron-mediated thermal boundary conductance for metal-metal interfaces [15,16]. This similitude is resulting from the fact that Eq. (5) is a generic description for properties of a two component system, where there is a mutual detrimental influence on the respective property of the other compound [17].

Expressing the ratio thickness as $n = d_1/d_2$, Eq. (7) can be rewritten as:

$$R_{12} = \frac{1}{2}\frac{n}{(n+1)}d_2 C \quad \text{or} \qquad (8)$$

$$R_{12} = \frac{1}{2}\frac{n}{(n+1)^2}LC \qquad (9)$$

From Eq. (8) it is clear that for a constant thickness $d_2$ and very thick $d_1$ layer, i.e., $d_1 \gg d_2$ and $n \gg 1$, the value of TBR is constant and approaches to $R_{12} = 1/2\ d_2\ C$. On the other hand, Eq. (9) shows that for a constant $L$ and $d_2 \to 0$, i.e., a pure single bulk system, the TBR goes to zero. Therefore, in both limits we can recover the expected values of the TBR.

Finally, the thermal resistivity of the SL can be expressed in terms of Norbury's rule given by:

$$\begin{aligned}\rho_{SL} &= \frac{d_1}{L}\rho_{1,bulk} + \frac{d_2}{L}\rho_{2,bulk} + 2\frac{R_{12}}{L} \\ &= x\rho_{1,bulk} + (1-x)\rho_{2,bulk} + x(1-x)C \text{ or} \\ \kappa_{SL} &= \left(\frac{x}{\kappa_{1,bulk}} + \frac{(1-x)}{\kappa_{2,bulk}} + x(1-x)C\right)^{-1}\end{aligned} \quad (10)$$

Using this simplistic fit, we will compare the dependence of the thermal conductivity on the digital alloy composition ($x = d_1/L$) of several SLs. All the experimental data discussed here were measured at room temperature, some of them were extracted from literature and other from our own measurements. In addition, the comparison with molecular dynamics simulations is also performed. The good agreement between the experiments and the Norbury-fit points out that this simplistic and analytical approach can capture some fundamental aspects of the behaviour of $\kappa$ and the TBR of NSSLs. The experimental data shown in this work, besides our results, were reproduced from the digitalization of the images and making an average of the experimental data (for the case of measurements with more than one technique) and the corresponding experimental uncertainties.

As first example, we examined the thermal conductivity and the TBR of (Si)$_{d1}$:(Ge)$_{d2}$ NSSLs. The experiments were performed in several SLs with a fixed layer-thickness ($d_1$ or $d_2$) and varying the thickness of the other layer ($d_2$ or $d_1$, respectively). The total number of periods, $N$, was kept constant $N = 21$. Full description of the experiments and the used methodology can be found in the work of Chen et al. [18] and references therein.

Figure 1a shows $\kappa_{SL}$ of (Si)$_{d1}$:(Ge)$_{0.41nm}$ (orange squares) and (Si)$_{6nm}$:(Ge)$_{d2}$ (green dots) as function of the digital alloy composition, $x = d_1/(d_1 + d_2)$. The $\kappa_{SL}$ was calculated from the best fit of the Eq. (10) and using fixed values of $\kappa_1 = 150$ [W (K m)$^{-1}$] and $\kappa_2 = 60$ [W (K m)$^{-1}$], corresponding to the room-temperature thermal conductivity values of Si and Ge bulk systems, respectively [19]. The best fit was estimated for $C = 3.7$ [W$^{-1}$ K m], with a 95% confidence bounds ranging $3.0 < C < 4.4$ [W$^{-1}$ K m] as it is shown with light-blue shadow in the same figure. The limit of bulk Si$_{(1-x)}$Ge$_x$ alloy is also plotted (black dashed-line) for comparison. It is worth noting the excellent agreement between the experimental data (green and orange dots) and Norbury-model (blue solid-line). In a similar manner, the experimental value of TBR was estimated using Eq. (3) and it was fitted using Eq. (7). The $C$-parameter and the corresponding confidence bounds were taken from the best fit of $\kappa_{SL}$. The linear dependence of TBR with the reduced thickness of the layers, $(1/d_1 + 1/d_2)^{-1}$, is shown in Figure 1b.

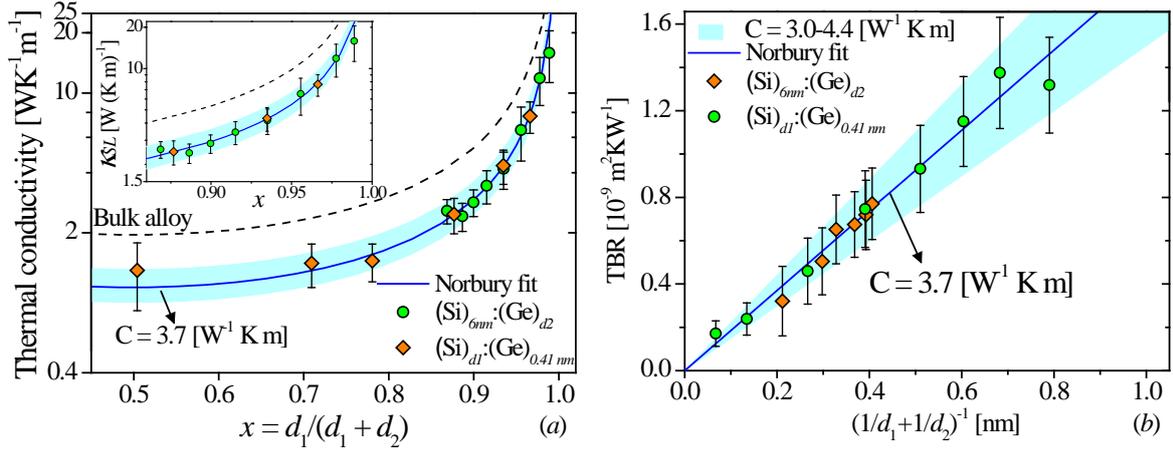

**Figure 1** *(a) Thermal conductivity of $(Si)_{d1}:(Ge)_{0.41nm}$ (orange squares) and $(Si)_{6nm}:(Ge)_{d2}$ (green dots) NSSLs as a function of the digital alloy composition, $x = d_1/(d_1+d_2)$. The blue solid-line shows the best fit of the C-parameter ($C = 3.7$ [W$^{-1}$ K m]) with fixed thermal conductivity values of $\kappa_1 = 150$ [W (K m)$^{-1}$] and $\kappa_2 = 60$ [W (K m)$^{-1}$], which correspond to the room-temperature bulk values of Si and Ge, respectively [19]. The light-blue shadow represents small variation of C in 95% range of confidence, $3.0 < C < 4.4$ [W$^{-1}$ K m]. The black dashed-line is the calculated alloy-bulk limit. (inset) Zoom around $x = 1$ for a better visualization of experimental and theoretical curves.*
*(b) Thermal boundary resistance of $(Si)_{d1}:(Ge)_{0.41nm}$ (orange squares) and $(Si)_{6nm}:(Ge)_{d2}$ (green dots) NSSLs as function of reduced layer thickness, $(1/d_1+1/d_2)^{-1}$. The blue solid-line represents the best fit of experimental data and Eq. (7), i.e. $C = 3.7$ [W$^{-1}$ K m], small variation around this value idem (a). All the measurements were taken from Ref. [18].*

The experimental data discussed in the second example, presented in Figure 2, were reported by Koh et al. [2]. In this case the thermal conductivity of several $(GaN)_{d1}:(AlN)_{4nm}$ SLs was measured by varying the thickness of the GaN, $d_1$, in the range of $2 < d_1 < 1000$ nm. Similar to the first example, the experimental TBR was evaluated based on Eq. (3) using the bulk values of GaN, $\kappa_1 = 195$ [W (K m)$^{-1}$], and AlN, $\kappa_2 = 319$ [W (K m)$^{-1}$], respectively [11]. The experimental TBR (green dots) as a function of the ratio of materials composition, $n = d_1/d_2$, is shown in Figure 2b. The theoretical curves were estimated from the best fit of Eq. (10) with the experimental data (see Figure 2a). The best fitting curves for $C = 0.72$ [W$^{-1}$ K m] and $C = 0.5$ [W$^{-1}$ K m] are traced with a green-solid and black-dotted lines, respectively. It is clear that for a very thick $d_1$ and constant $d_2$ thickness layer, the TBR tends to the constant value given by the Eq. (8), $R_{12} = 1/2\ d_2\ C$ (see the black short-dotted line of Figure 2b).

The experimental thermal conductivity, green dots, as a function of digital alloy composition is displayed Figure 2a. The corresponding best fitting curves for $C = 0.72$ and $0.5$ [W$^{-1}$ K m] as well as the alloy bulk limit are shown in green-solid, black-dotted and black-dashed lines, respectively. The light green shadow, in Figure 2a and b, represents 95% range of confidence of the best fitting value of the C-parameter.

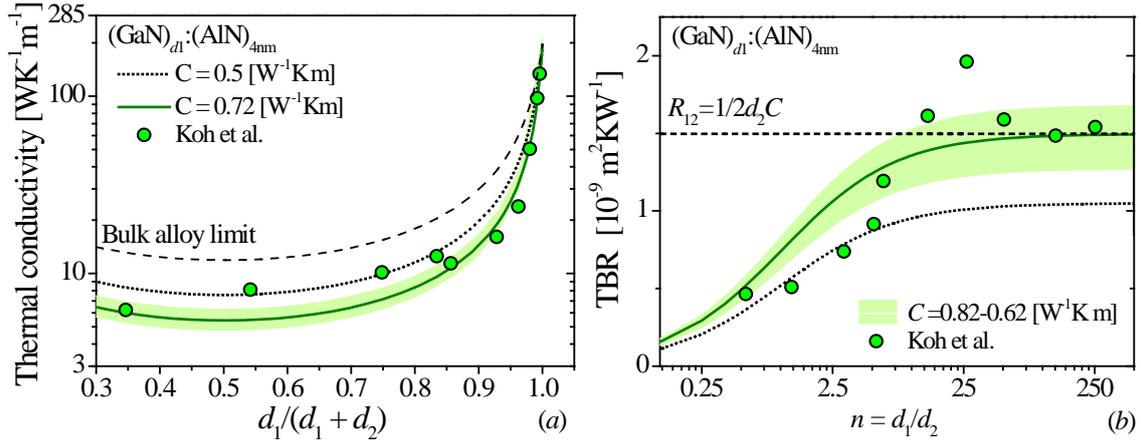

**Figure 2** *(a) Experimental (green dots) and fitted (green solid- and black dotted-lines) thermal conductivity of $(GaN)_{d1}$:$(AlN)_{4nm}$ as a function of the digital alloy composition, $x = d_1/(d_1+d_2)$. Two different values of C-parameter are shown by a green solid-line (C = 0.72 [W$^{-1}$ K m]) and a dash-dotted-black line (C = 0.5 [W$^{-1}$ K m]). (b) Experimental (green dots) and fitted (solid-green and dash-dotted black line) thermal boundary resistance as function of the ratio of layer thickness $(d_1/d_2)$ of the same set of samples presented in (a). The black short-dotted line represents the limit for very thick $d_1$ layer $R_{12} = 1/2\, d_2\, C$, as it was discussed (see Eq. (7)). The light green-shadow represents 95% range of confidence of the C-parameter, valid for both figures. All the experimental data were taken from Ref.* [2].

A third example is presented for $(TiNiSn)_{d1}$:$(HfNiSn)_{d2}$. In this experiment the period thickness was kept constant, $L \sim 3.6$ nm, and the ratio of the layers, $n$, was varied in a range of $0.2 < n < 5$. The thickness of the layers varied linearly as $d_1 = 6 - m$ and $d_2 = m$, with $1 < m < 5$ representing the number of units cells. All the measurements were performed in our laboratory using the standard three-omega method. A detailed description of the growth conditions and thermal characterization can be found in Refs. 8, 17, 20. The TBR was estimated from Eq. (3) using the measured thermal conductivity for pure TiNiSn ($\kappa_1 = 3.96$ [W (K m)$^{-1}$]) and HfNiSn ($\kappa_1 = 1.59$ [W (K m)$^{-1}$]) 1 μm thin films.

The TBR as a function of $n$ with a constant $L$ is displayed in Figure 3b. Note that for a constant period thickness $L$ and $d_2 \to 0$, the TBR approaches to zero, as it was shown by the Eq. (9). Additionally, the period-thickness dependence of TBR of a different symmetrical series, $d_1 = d_2$, is also plotted in Figure 3c. The experimental TBR of (TiNiSn):(HfNiSn) [8] and $(Ti_{0.7}W_{0.3}N)$:$(Al_{0.72}Sc_{0.28}N)$ [3] SLs are displayed with cyan dots and orange squares, respectively. The latest SL is just plotted as an example and all the experimental details can be found in the work of Saha et al. in Ref. [3]. For the symmetrical case, we expected a deviation of the linear dependence with period thickness because our model is based on the consideration of $d_1 \neq d_2$. However, this linear dependence of TBR for small period thickness gives the same fluctuation range of the *C*-parameter values in concordance with non-symmetric case. Similar dependence have been also observed in other theoretical reports [21–23]. The dependence of the $\kappa_{SL}$ for NSSLs is presented in Figure 3a, showing a good agreement with the Norbury`s rule (blue-solid line).

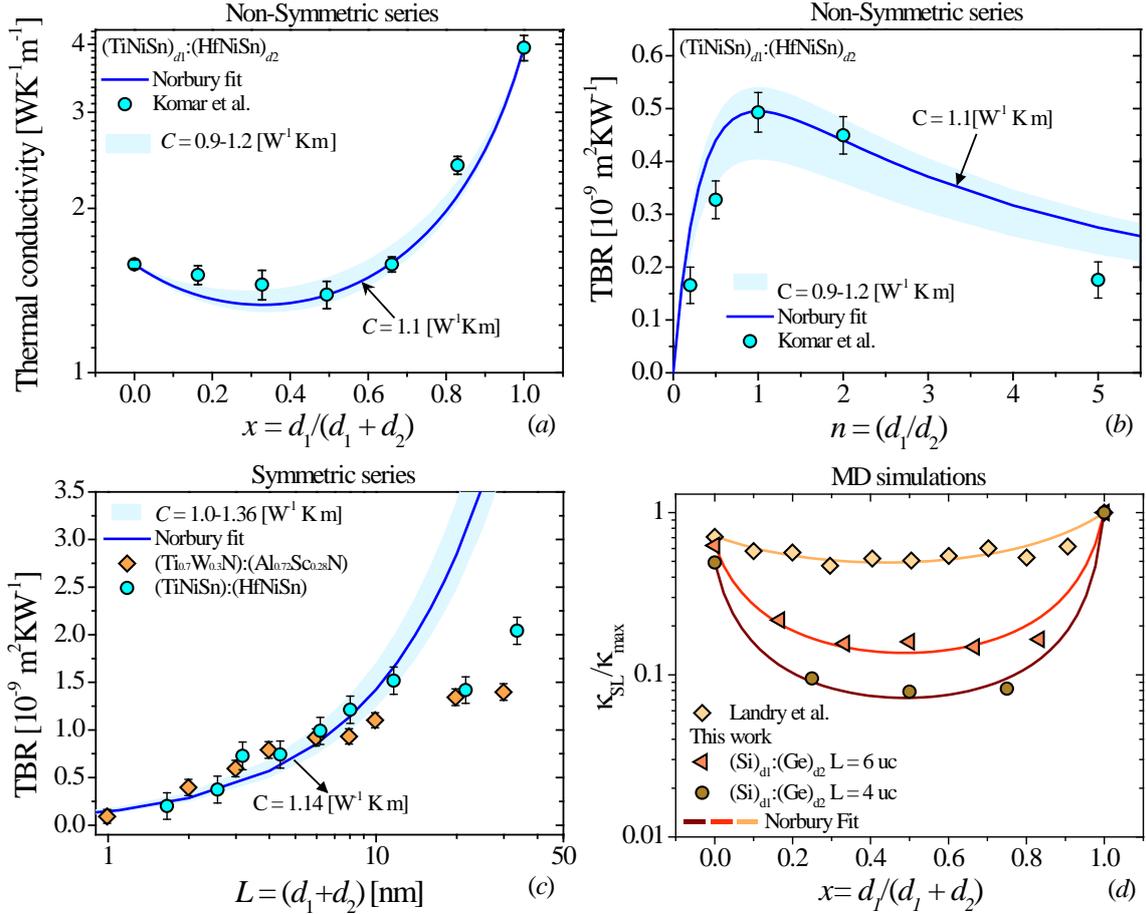

**Figure 3** *(a) Thermal conductivity of non-symmetric series of $(TiNiSn)_{d1}:(HfNiSn)_{d2}$ SLs as a function of the digital alloy composition, $x = d_1/(d_1+d_2)$. (b-c) Thermal boundary resistance of $(TiNiSn)_{d1}:(HfNiSn)_{d2}$, cyan dots, as a function of: (b) ratio of material composition with a constant period* [17], *i.e., $d_1 \neq d_2$; and (c) period thickness with a constant ratio of the material composition* [8], *i.e., $d_1 = d_2$. As an example, the TBR of $(Ti_{0.7}W_{0.3}N):(Al_{0.72}Sc_{0.28}N)$ (orange squares) is also included in (c). The experimental details can be found in work of Saha et al. see Ref.* [3]. *(d) Molecular dynamic (MD) calculations of the thermal conductivity of Ar-based (yellow squared,* [24]*) and two (Si):(Ge) (with a period length of 6 and 4 uc, orange triangles and brown circles, respectively) NSSLs. The thermal conductivity has been normalized using the maximum thermal conductivity for the better visualization $\kappa_{SL}/\kappa_{max}$.*

Based on the previous examples, it is clear that the interface TBR strongly depends on period thickness as well as the ratio of the material composition. Similar observations were found by Saha et al. [3] in $(Ti_{0.7}W_{0.3}N):(Al_{0.72}Sc_{0.28}N)$ metal/semiconductors symmetric-SLs. As it is shown in Figure 3c, a linear dependence of the TBR with the period thickness was found for small period thickness (note that the *x*-axis is in logarithmic scale). Saha and co-workers associated this dependence to contribution of the coherent phonons (band-folded ones) to the total thermal conductivity.

Finally, a fourth example is also presented a comparison with molecular dynamic simulations (see Figure 3d). In this case we compared the Norbury's rule with a reported MD simulations of the $\kappa$ of Ar-based face-centered-cubic lattice [24] and two (Si):(Ge) NSSL. The thermal conductivity of the latest SLs were calculated by using nanoMaterials nanoscale heat transport online software [25,26]. Here, the period length was kept constant ($L = 6$ and 4 uc, orange triangles and brown circles, respectively) and the thickness of the individual layers was linearly varied as $d_1 = L - m$ and $d_2 = m$, with $1 < m < L$ the number of units cells. The total number of periods was kept constant with $N = 4$ and 5, respectively.

As we can see in the Figure 3d, a similar quadratic shape of the $\kappa$ is also found in the MD simulations matching quite well Norbury's rule. This last example makes evident that the modification of Norbury's rule is a generic description for properties of a two component system, where there is a mutual detrimental influence on the respective property of the other compound.

It is natural to think that in some extreme scenarios such as ballistic-ballistic or ballistic diffusive transport through the SL the Norbury's fit ansatz should break down. Nevertheless, the model still fits quite well. For example in the Figure 1a one of the SLs has a composition of $(Si)_{12nm}:(Ge)_{0.42nm}$ (~ 97% Si (24 layers) and 3% Ge (1 layer)). In such an extreme case, it is likely that the Ge layers will reduce the mean free paths of the phonons primarily in Si and we will get a substantial reduction in Si thermal conductivity [27], far below what we are using to fit this model. However, in such extreme scenario the thermal conductivity of the whole SL is completely dominated by the quadratic term of the Norbury's fit, and the role of the linear elements is completely negligible. The other extreme scenario is presented in the Figure 2a. Here Koh et al. [2] have grown samples with a composition of $(GaN)_{1012nm}:(AlN)_{4nm}$ (~ 99.6 % GaN and 0.4% of AlN). In this extreme the linear elements play a role and the fit cannot be improved by changing the values of the $C$-parameter.

The form of Eq. (10) captures that small interactions between the systems have a linear behaviour, whereas close to the minimum a quadratic behaviour is observed even in atomically flat interfaces such is the case of the MD simulations.

**Conclusions**

An extension of the Norbury's rule for the thermal conductivity calculation of $(A)_{d1}:(B)_{d2}$ non-symmetrical superlattices was presented. Utilizing this phenomenological description, the dependence of the interface thermal resistance with period thickness and the ratio of the material composition was found. The outstanding agreement between experimental data and the Norbury rule offers a simplistic and quick view of the impact of period thickness ($L$) and ratio of material composition ($n$) on $\kappa_{SL}$ and TBR. However, the successful fitting with the Norbury's rule model will not guarantee that the physical mechanism is captured, as clearly the superlattices discussed here are not alloys and also the description of the superlattices as digital alloys is an oversimplification. Nevertheless the current approach is a simple and easy to implement method that captures the behaviour of the thermal conductivity of SLs.

**Acknowledgments**

We gratefully acknowledge financial support by the Deutsche Forschungsgemeinschaft, DFG, Germany, [Grants No. Ja821/4 within SPP 1386 (Nanostructured Thermoelectric Materials) and No. Ja821/7-1 within SPP 1538 (Spin Caloric Transport)] and the Graduate School of Excellence Material Science in Mainz (GSC 266). We thank J. Jaramillo-Fernandez and J. Ordonez for the valuable discussion.

**References**

[1] E.T. Swartz, and R.O. Pohl, Thermal boundary resistance. *Reviews of Modern Physics*, Vol. 61, No. 3, pp. 605–668, 1989.

[2] Y.K. Koh, Y. Cao, D.G. Cahill, et al. Heat-transport mechanisms in superlattices, *Advanced Functional Materials*, Vol. 19, No. 4, pp. 610–615, 2009.

[3] B. Saha, Y.R. Koh, J. Comparan, et al., Cross-plane thermal conductivity of (Ti,W)N/(Al,Sc)N metal/semiconductor superlattices, *Physical Review B*, Vol. 93, No. 4, pp. 045311, 2016.

[4] P. E. Hopkins, Thermal transport across solid interfaces with nanoscale imperfections: effects of roughness, disorder, dislocations, and bonding on thermal boundary conductance, *ISRN Mechanical Engineering*, Vol. 2013, pp. 1–19, 2013.

[5] R. Venkatasubramanian, Lattice thermal conductivity reduction and phonon localization like behavior in superlattice structures, *Physical Review B*, Vol. 61, No. 4, pp. 3091–3097, 2000.

[6] S. Chakraborty, C. A. Kleint, A. Heinrich, et al., Thermal conductivity in strain symmetrized Si/Ge superlattices on Si(111), *Applied Physics Letters*, Vol. 83, No. 20, pp. 4184, 2003.

[7] J. Ravichandran, A. K. Yadav, R. Cheaito, et al. Crossover from incoherent to coherent phonon scattering in epitaxial oxide superlattices, *Nature Materials*, Vol. 13, No. 2, pp. 168–172, 2013.

[8] P. Hołuj, C. Euler, B. Balke, et al., Reduced thermal conductivity of TiNiSn/HfNiSn superlattices, *Physical Review B*, Vol. 92, No. 12, pp. 125436, 2015.

[9] M. V. Simkin, and G. D. Mahan, Minimum thermal conductivity of superlattices, *Physical Review Letters*, Vol. 84, No. 5, pp. 927–930, 2000.

[10] S. Adachi, Lattice thermal resistivity of III–V compound alloys, *Journal of Applied Physics*, Vol. 54, No. 4, pp. 1844, 1983.

[11] S. Adachi, Lattice thermal conductivity of group-IV and III–V semiconductor alloys, *Journal of Applied Physics*, Vol. 102, No. 6, pp. 063502, 2007.

[12] M. E. Levinshtein, S. L. Rumyantsev, and M. S. Shur, 2001. Properties of advanced semiconductor materials: GaN, AlN, InN, BN, SiC, SiGe, UK: John Wiley & Sons.

[13] A. L. Norbury, The electrical resistivity of dilute metallic solid solutions, *Transactions of the Faraday Society*, Vol. 16, pp. 570, 1921.

[14] B. Abeles, Lattice thermal conductivity of disordered semiconductor alloys at high temperatures, *Physical Review*, Vol. 131, No. 5, pp. 1906–1911, 1963.

[15] B. C. Gundrum, D. G. Cahill, and R. S. Averback. Thermal conductance of metal-metal interfaces, *Physical Review B*, Vol. 72, No. 24, pp. 245426, 2005.

[16] P. E. Hopkins, T. E. Beechem, J. C. Duda, et al. Effects of subconduction band excitations on thermal conductance at metal-metal interfaces, *Applied Physics Letters*, Vol. 96, No. 1, 011907, 2010.

[17] P. Komar, E. Chávez-Ángel, C. Euler, et al. Tailoring of the electrical and thermal properties using ultra-short period non-symmetric superlattices, *APL Materials*, Vol. 4, No. 10, pp. 104902, 2016.

[18] P. Chen, N. A. Katcho, J. P. Feser, et al. Role of surface-segregation-driven intermixing on the thermal transport through planar Si/Ge superlattices. *Physical Review Letters*, Vol. 111, No. 11, pp. 115901, 2013.

[19] C. J. Glassbrenner and G. A. Slack. Thermal conductivity of silicon and germanium from 3°K to the melting point, *Physical Review*, Vol. 134, pp. 1058–1069, 1964.

[20] P. Komar, T. Jaeger, C. Euler, et al. Half-Heusler superlattices as model systems for nanostructured thermoelectrics, *Physica Status Solidi (a)*, Vol. 213, pp. 732–738, 2015.

[21] Y. Chalopin, K. Esfarjani, A. Henry, et al. Thermal interface conductance in Si/Ge superlattices by equilibrium molecular dynamics, *Physical Review B*, Vol. 85, No. 19, 195302, 2012.

[22] Y. Ni, Y. Chalopin, and S. Volz. Significant thickness dependence of the thermal resistance between few-layer graphenes, *Applied Physics Letters*, Vol. 103, No. 6, pp. 061906, 2013.

[23] Z. Liang, K. Sasikumar, and P. Keblinski. Thermal transport across a substrate-thin-film interface: effects of film thickness and surface roughness. *Physical Review Letters*, Vol. 113, No. 6, pp. 065901, 2014.

[24] E. S. Landry, M. I. Hussein, and A. J. H. McGaughey, Complex superlattice unit cell designs for reduced thermal conductivity. *Physical Review B*, Vol. 77, No. 18, pp. 184302, 2008.

[25] K. H. Lin, and A. Stracha. Thermal transport in SiGe superlattice thin films and nanowires: effects of specimen and periodic lengths, *Physical Review B*, Vol. 87, pp. 115302, 2013.

[26] K. H. Lin, S. Sullivan, M. J. Cherukara, et al. *nanoMaterials nanoscale heat transport*, 2016. Available at: https://nanohub.org/resources/nmstthermal

[27] E. Chávez-Ángel, J.S. Reparaz, J. Gomis-Bresco, et al. Reduction of the thermal conductivity in free-standing silicon nano-membranes investigated by non-invasive Raman thermometry. *APL Materials*. Vol. 2, No. 1, pp. 012113, 2014.